\documentclass[twoside,reqno]{HERON}
\usepackage{epsfig,cite,colordvi}
\usepackage{graphicx}
\usepackage{url}
\usepackage{amssymb,amsmath,amscd,epsf}
\usepackage{times}
\usepackage{makeidx}
\def\dblone{\hbox{$1\hskip -1.2pt\vrule depth 0pt height 1.6ex width 0.7pt
     \vrule depth 0pt height 0.3pt width 0.12em$}}

\begin{document}

\title{Microscopic  Optical Potentials from Chiral Forces and Ab Initio Nuclear Densities}

\runningheads{Microscopic  Optical Potentials from Chiral Forces and Ab Initio Nuclear Densities}{C. Giusti, M. Vorabbi, P. Finelli}

\begin{start}

\author{C. Giusti}{1}, \coauthor{M. Vorabbi}{2}, \coauthor{P. Finelli}{3}

\index{Giusti, C.}
\index{Vorabbi, M.}
\index{Finelli, P.}

\address{INFN, Sezione di Pavia,  Via A. Bassi 6, I-27100 Pavia, Italy}{1}

\address{Department of Physics, University of Surrey, Guildford, GU2 7XH, UK}{2}

\address{Dipartimento di Fisica e Astronomia, Universit\`{a} degli Studi di Bologna and INFN, Sezione di Bologna, Via Irnerio 46, I-40126 Bologna, Italy}{3},

\begin{Abstract}
We derived microscopic optical potentials (OPs) for elastic nucleon-nucleus scattering within the framework of  chiral effective field theories at the first-order term of the spectator expansion of the Watson multiple-scattering theory and adopting the impulse approximation. Our OPs are
derived by folding \emph {ab initio} nuclear densities with a nucleon-nucleon ($NN$) $t$ matrix computed with a consistent chiral interaction. The results of our OPs are in  good agreement with the experimental data. Recent achievements of our work are reviewed in this contribution.
\end{Abstract}
\end{start}

\section{Introduction}

The optical potential (OP) provides a successful tool to describe elastic nucleon- nucleus($NA$) scattering. Its use can be extended to inelastic scattering and to the calculation of the cross section of a wide variety of nuclear reactions. The basic idea is to describe the $NA$ interaction with an effective complex and energy-dependent potential \cite{FESHBACH1958357,hodgson1963}.
The imaginary part accounts for the flux lost from the elastic channel to open inelastic and reaction channels, while the energy dependence and nonlocalities account for the underlying many-nucleon dynamics.

Phenomenological and microscopic approaches have been used to derive an OP. Phenomenological OPs are obtained assuming an analytical form and a dependence on a number of adjustable parameters for the real and imaginary parts that characterize the shape of the nuclear density distribution and that vary with the nucleon energy and the nuclear mass number. The values of the parameters are determined through a fit to elastic $pA$ scattering data. Global OPs, available for a wide range of nuclei and energies \cite{Hebborn,KD}, are quite successful in the description of elastic scattering data and are usually adopted for the calculation of the cross section of many nuclear reactions. 
Microscopic OPs are the result of a microscopic calculation and not of a fitting procedure and are, therefore, more theoretically founded, but in principle require the solution of the full many-body problem for the incident nucleon and all the nucleons of the target nucleus, which is a tremendous task, often beyond current computing capabilities. Some approximations are needed to reduce the problem to a tractable form and the reliability of the OP depends on the
reliability of the adopted approximations. In general, one would expect that a microscopic OP can be less able to describe elastic $NA$ scattering data than a
phenomenological OP, 
but it can have a greater predictive power when applied to situations for which experimental data are not yet available.

We believe that the derivation of 
a microscopic OP, starting from $NN$ and three-nucleon (3N) interactions, where the approximations and uncertainties of the model are reduced as much as possible, 
is mandatory to 
provide reliable predictions for a wide range of nuclei. This is particularly important for nuclei away from stability, whose study represents a frontier in nuclear science over the coming years and which will be probed at new rare-isotope beam facilities worldwide \cite{Hebborn}.

In a series of papers over the last years \cite{Vorabbi1,Vorabbi2,Vorabbi3,Vorabbi4,Vorabbi5,Vorabbi6,Vorabbi7,Vorabbi8} we derived microscopic OPs for elastic (anti)nucleon-nucleus scattering from chiral nuclear interactions. Our OPs have been obtained at the first-order term of the spectator expansion of the Watson multiple-scattering theory \cite{Watson,KMT} and adopting the impulse approximation (IA). 
The idea was to start from a relatively simple model and with subsequent steps  improve and extend the model. 

An overview of the latest achievements of our work is presented in this contribution. 
In Section~2 we outline the theoretical framework used to calculate our microscopic OPs. Our latest achievements and their main findings are discussed in Section~3. Our conclusions and perspectives are drawn in Section~4.

 \section{Theoretical framework}

In this section we outline only the main steps of the derivation of our microscopic OPs. More details can be found in  \cite{Vorabbi1,Vorabbi2,Vorabbi3,Vorabbi4,Vorabbi5,Vorabbi6,Vorabbi7,Vorabbi8}.  

The standard approach to the elastic scattering of a nucleon from a target nucleus  of $A$ particles is the separation of the full Lippmann-Schwinger (LS) equation for the transition operator 
\begin{equation}
\label{generalscatteq}
T = V \left( 1 + G_0 (E) T \right)\,  
\end{equation}
into two parts, \textit{i.e.} an integral equation for $T$
\begin{equation}\label{firsttamp}
T = U \left(1+ G_0 (E) P T\right) \, ,
\end{equation}
where $U$ is the optical potential operator, and an integral
equation for $U$
\begin{equation}\label{optpoteq}
U = V \left(1+ G_0 (E) Q U \right)\, .
\end{equation}

In the above equations $V$ is the external interaction, $G_0 (E)$ the free Green's function for the $(A+1)$-nucleon system, and $P$ and $Q = \dblone-P$  projection operators that select the elastic channel.

A consistent framework to compute $U$ and $T$ is provided by the spectator expansion, that is based on the multiple-scattering theory \cite{Watson}. We retain only the first-order term, corresponding to the single-scattering approximation, where only one target-nucleon interacts with the projectile. Moreover, we adopt the IA, where nuclear binding on the interacting target nucleon is neglected~\cite{Vorabbi1}. The adopted approximations reduce the complexity of the original many-body problem to a form where, in practice, we have to solve only two-body equations.

After some manipulations, the OP is obtained as  a folding integral of the two main ingredients of the model: the target density and the $NN$ $t$ matrix, as
\begin{equation}\label{fullfoldingop}
\begin{split}
U ({\bf q},{\bf K}; E ) = \sum_{N=p,n} &\int d{\bf P} \, \eta ({\bf q},{\bf K},{\bf P}) \,
t_{NN} \\
&\times \left[ {\bf q} , \frac{1}{2} \left( \frac{A+1}{A} {\bf K} + \sqrt{\frac{A-1}{A}} {\bf P} \right);E \right] \\
&\times \rho_N \left( {\bf P} + \sqrt{\frac{A-1}{A}} \frac{{\bf q}}{2} , {\bf P} - \sqrt{\frac{A-1}{A}} \frac{{\bf q}}{2} \right) \, ,
\end{split}
\end{equation}
where ${\bf q}$ and ${\bf K}$ represent the momentum transfer and the average momentum, respectively. Here ${\bf P}$ is an integration variable, $t_{NN}$ is the $NN$
$t$ matrix and $\rho_N$ is the one-body nuclear density matrix. 
The parameter $\eta$ is the M\"{o}ller factor, that imposes the Lorentz invariance of the flux when we pass from
the $NA$ to the $NN$ frame in which the $t$ matrices are evaluated, and $E$ is the energy at which the $t$ matrices are evaluated. 

In our first papers \cite{Vorabbi1,Vorabbi2,Vorabbi3} the use of local neutron and proton densities from a relativistic mean-field model \cite{Nik1} gives the  OP in a factorized form, the so-called optimum factorization approximation, as the product of the density and the $t$ matrix, thus avoiding the calculation of the folding integral of Eq.~(\ref{fullfoldingop}). 
For the $NN$ interaction in $t_{NN}$ we used two versions of chiral potentials at fourth order (N$^3$LO) in the chiral expansion \cite{EM,EGM} in Ref. \cite{Vorabbi1} and at fifth order (N$^4$LO) \cite{EKM,EMN} in Ref. \cite{Vorabbi2}. We studied the chiral convergence of the potentials in reproducing elastic $pA$ scattering data. The results show that it is mandatory to use chiral potentials at least at N$^3$LO. Lower-order potentials are unable to describe the shape and the magnitude of the 
scattering observables
of elastic $pA$ scattering. The results obtained with chiral potentials at N$^4$LO are neither better nor worse then those obtained with chiral potentials at N$^3$LO. In Ref. \cite{Vorabbi3} we compared the performances of our OPs and those of a successful phenomenological OP \cite{KD} in the description of the experimental data over a wide range of nuclei, including isotopic chains, in  a proton-energy range between 150 and 330 MeV. The agreement of our OPs with the data is sometimes worse and sometimes better, but overall comparable to the agreement given by the phenomenological OP, in particular, it is better for energies close and above 200 MeV. 

The OP model was improved in Ref. \cite{Vorabbi4}, where the
folding integral of $t_{NN}$ and a microscopic nonlocal density obtained with the \textit{ab initio} no-core shell model \cite{barrett} (NCSM) approach, utilizing $NN$ and $3N$ chiral interactions, was calculated.
The same chiral $NN$ interaction employed to calculate the
nuclear density is used to calculate $t_{NN}$. This guarantees the consistency of the theoretical framework and improves the soundness of the numerical predictions of the OP model. The same approach, with the same
nonlocal NCSM density, was extended to elastic scattering of antiprotons off several target nuclei \cite{Vorabbi5}. In the calculations of $t_{\bar{N}N}$ the first $\bar{N}N$ chiral interaction at N$^3$LO \cite{DHM} was used.
Our results are in good agreement with the existing experimental
data \cite{Vorabbi5}.

\section{Recent achievements}

The OPs of Ref. \cite{Vorabbi4} uses $NN$ and $3N$ interactions to
calculate the target density, the structure part of the OP, while the dynamic part, $t_{NN}$, includes only the $NN$ interaction. Even if we can argue that the impact of $3N$ forces is more important in the density, since reproducing the nuclear radii is essential for a proper description of the diffraction minima in the differential cross section,
a more consistent OP would require the use of the same $NN$ and $3N$
potentials both in the dynamic and in the structure parts. 
Unfortunately, the exact treatment of the $3N$ interaction is a very hard task that is beyond our present capabilities.

Many-nucleon forces can be divided into genuine contributions, arising from the nuclear Hamiltonian, and induced terms, coming from the process of solving the nuclear many-body problem.
Genuine contributions enter directly into the definition of the nuclear Hamiltonian in terms of the active degrees of freedom chosen to describe the nuclear systems.
Recently, with a suitable approximation, we have 
investigated the role of genuine $3N$ forces in the dynamic part of the OP already at the level of the single-scattering approximation between the projectile and the target nucleon \cite{Vorabbi6}. The pure $3N$ force is approximated by a density-dependent $NN$ force, obtained by averaging the third nucleon momenta over the Fermi sphere, that is added as a medium correction of the bare $NN$ force used to calculate the $t$ matrix. We constructed the density-dependent $NN$ force following the procedure proposed in Ref. \cite{Holt}.

\begin{figure}[ht]
\begin{minipage}[h]{0.5\textwidth}
\includegraphics[width=5.2cm]{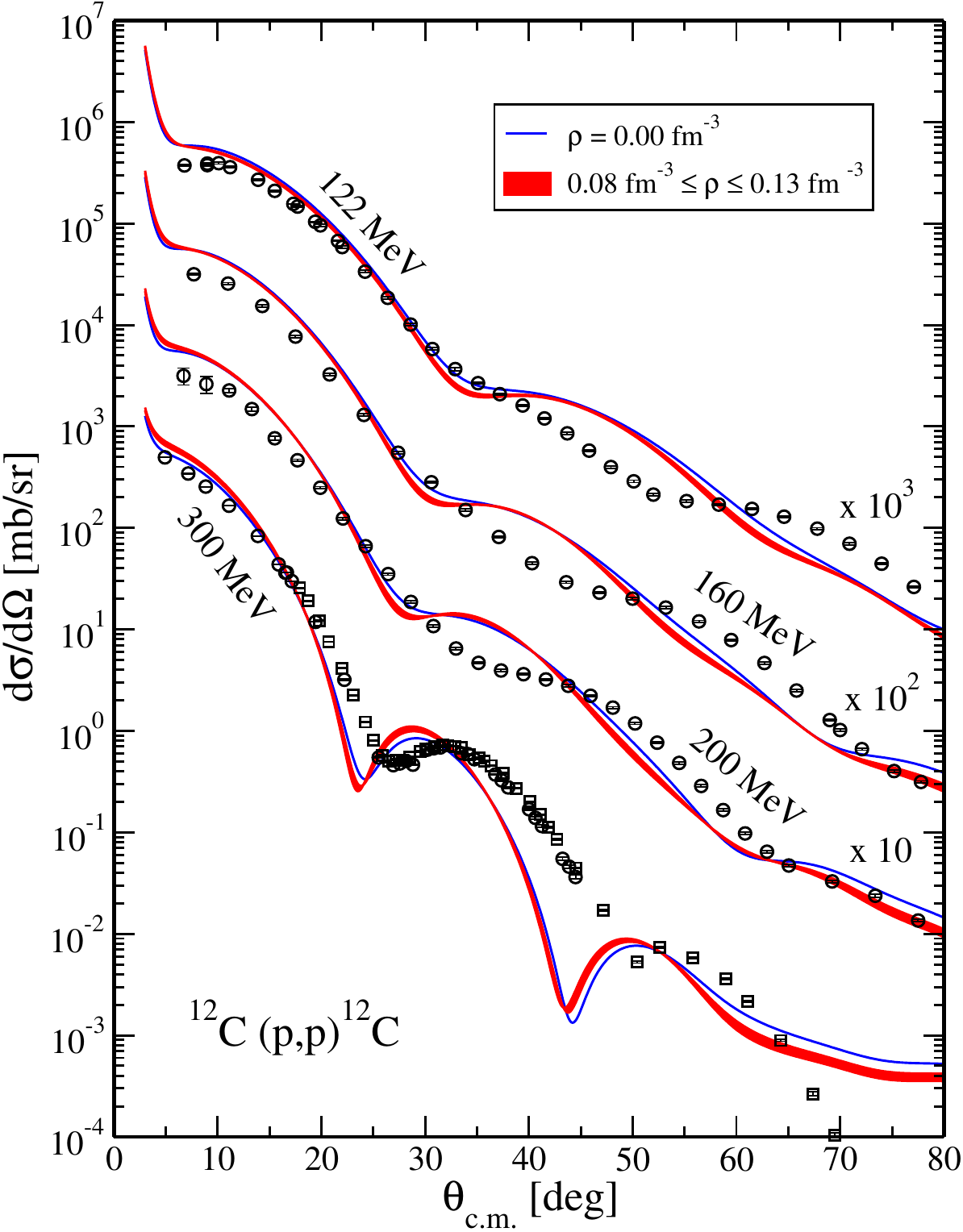}
\end{minipage} \hspace{0.01\textwidth}
\begin{minipage}[h]{0.5\textwidth}
\includegraphics[width=5.2cm]{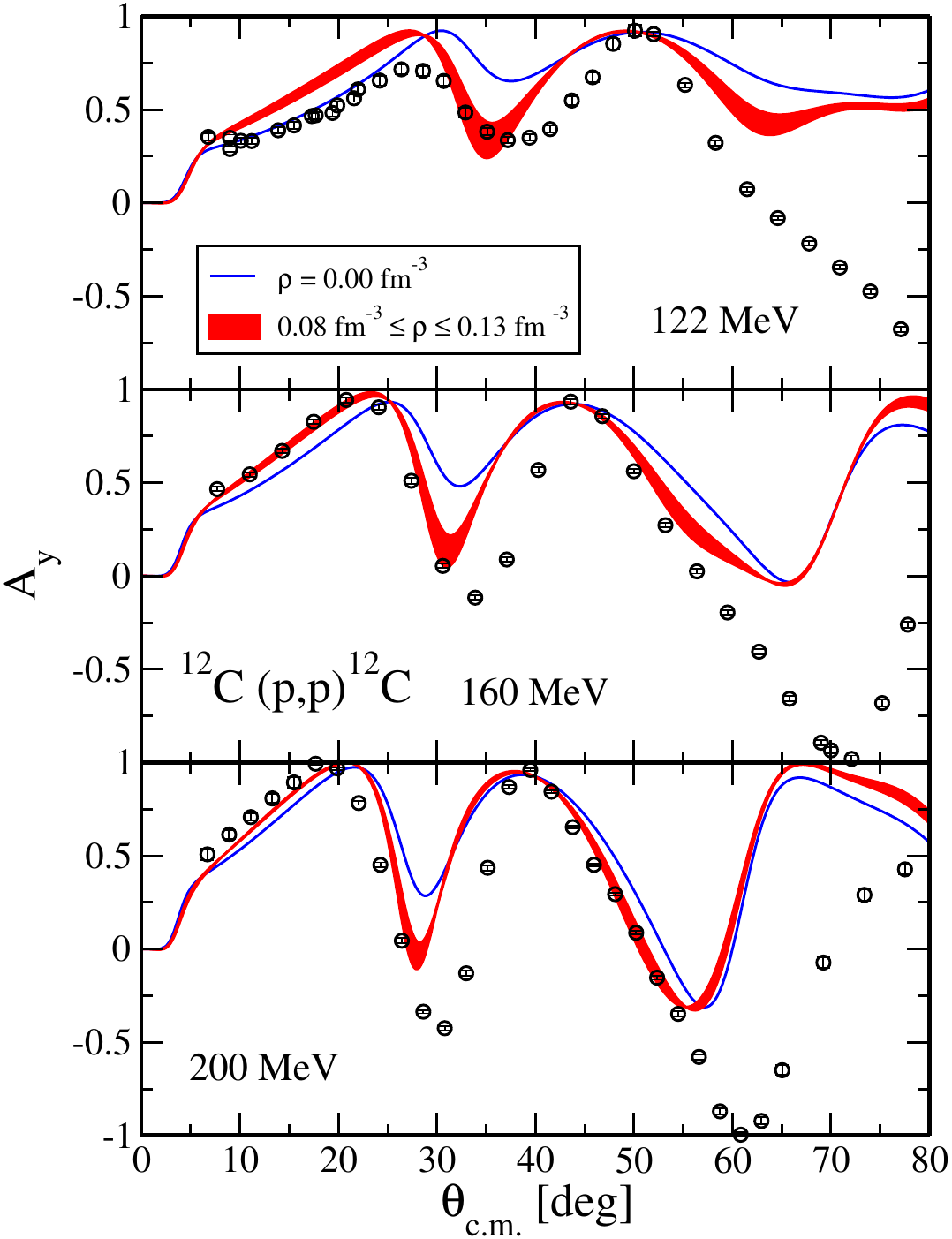}
\end{minipage}
\caption{Differential cross section (left panel) and analyzing power (right panel) as a function of the c.m. scattering angle   
for elastic proton scattering off $^{12}$C  at different laboratory energies. The red bands show the results obtained using Eq.(\ref{fullfoldingop}), where the $t_{pN}$ matrix is computed with the $pN$ chiral interaction of Ref. \cite{EMN}, supplemented by a density dependent $NN$ interaction where the matter density parameter $\rho$ has been varied between $0.08$ fm$^{-3}$ and $0.13$ fm$^{-3}$. The solid blue lines are obtained with $\rho= 0$ fm$^{-3}$. 
The one-body nonlocal density matrices are computed with the NCSM method using $NN$ \cite{EMN} and $3N$ \cite{Navratil2007,Gysbers2019} chiral interactions. Experimental data from Refs. 
\cite{PhysRevC.21,PhysRevC.27,PhysRevC.23,PhysRevC.81,PhysRevC.31}.}
\label{fig:3N}
\end{figure}

Even if the 3N force is treated in an approximate way, this method extends our previous OP model and allows a direct comparison of our present and previous results. A few examples of the impact of genuine $3N$ forces in the dynamic part of our OPs are shown in Figure~\ref{fig:3N}, where the calculated differential cross section and analyzing power ($A_y$), as a function of the center-of-mass (c.m.) scattering angle, are displayed for elastic proton scattering off $^{12}$C at  energies between 122 and  300 MeV and compared with the experimental data. All the results are obtained with the same one-body \textit{ab initio} density matrix from the NCSM approach using $NN$ and $3N$ chiral interactions. The red bands show the results obtained with $t_{pN}$ calculated with the $pN$ chiral interaction at N$^4$LO \cite{EMN} supplemented by a density-dependent $NN$ interaction where the matter density $\rho$ has been varied between reasonable values, going from surfacelike to bulklike densities. The blue lines correspond to $\rho=0$ fm$^{-3}$, \textit{i.e.} only the $pN$ interaction is considered in the calculation of  $t_{pN}$. 
The effects of genuine 3N forces turn out to be negligible for the cross section, where all curves basically overlap and are in reasonable agreement with the experimental data, and somewhat larger for $A_y$, where the $3N$ contribution improves the description of the empirical data. 

Our microscopic OP has been extended to nonzero spin nuclei \cite{Vorabbi7}. The extension requires some changes in the derivation of the OP and in the formalism. The main difference is that the density of a nonzero spin target displays an additional dependence on the initial and final third component of the spin which is then propagated to the OP and calculations get more and more involved and time consuming with the increasing value of the target spin.

\begin{figure}[ht]
\begin{minipage}[h]{0.5\textwidth}
\includegraphics[width=5.2cm]{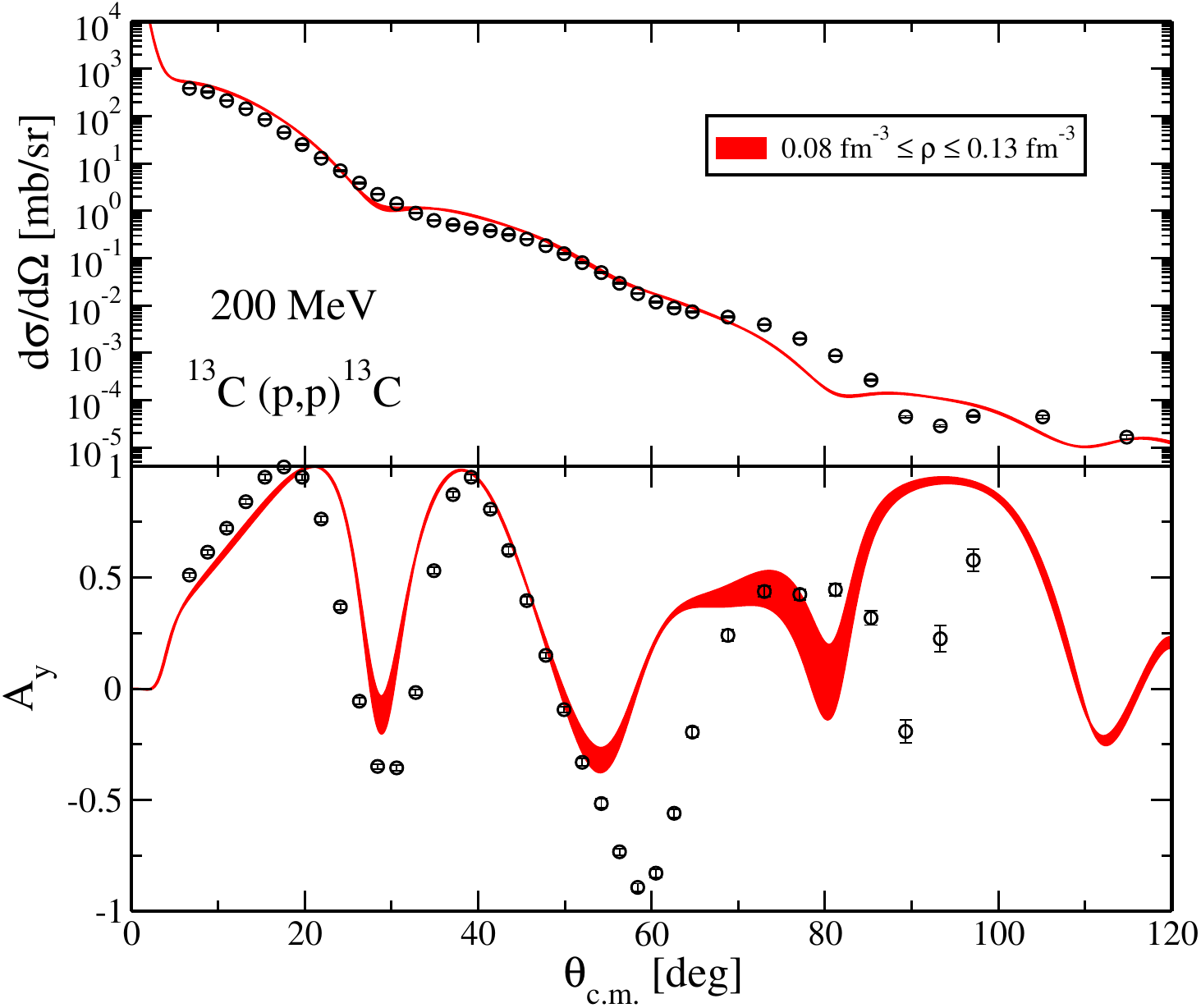}
\end{minipage} \hspace{0.01\textwidth}
\begin{minipage}[h]{0.5\textwidth}
\includegraphics[width=5.2cm]{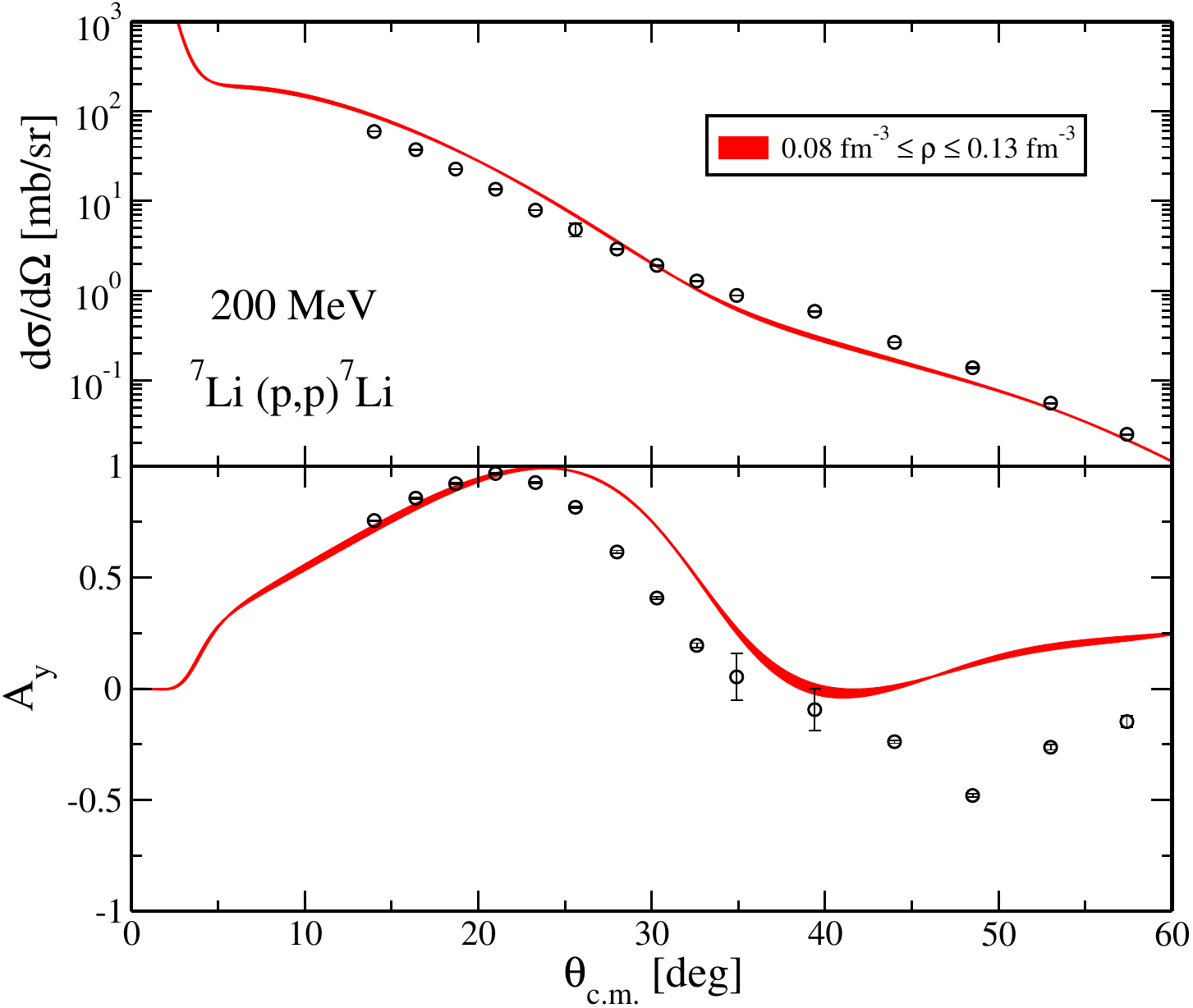}
\end{minipage}
\caption{Differential cross section (top panels) and analyzing power (bottom panels), as a function of the c.m. scattering angle, 
for elastic proton scattering off $^{13}$C, $J^\pi =1/2^-$, (left panels) and $^7$Li, $J^\pi =3/2^-$ (right panels), at  200 MeV laboratory energy. The bands have the same meaning as in in Figure~\ref{fig:3N}. The nuclear densities are computed with the NCSM method using $NN$ \cite{EMN} and $3N$ \cite{Navratil2007,Gysbers2019} chiral interactions. Experimental data from Refs. \cite{PhysRevC.23a,PhysRevC.43}.}
\label{fig:nonzero_spin}
\end{figure}

Calculations have been performed for the differential cross section and the analyzing power of elastic proton scattering off a set of nuclei with different values of the spin in their ground state, between $J =1/2$ and $3$,  and the results have been compared with the available data \cite{Vorabbi7}. A couple of examples are shown in Figure~\ref{fig:nonzero_spin}, for elastic proton scattering off $^{13}$C (with spin and parity quantum number $J^\pi =1/2^-$) and $^7$Li ($J^\pi =3/2^-$) at  200 MeV.
The effects of genuine $3N$ forces are small on the cross section and a bit larger on $A_y$ . The bands, indicating the differences due to different values of the matter density, are thin for the cross section a bit larger for $A_y$. The impact of $3N$ forces is comparable to what obtained for spin-zero targets. The agreement with data is generally satisfactory and of the same quality as in the case of spin-zero nuclei.

The use of the \textit{ab initio} NCSM method  for the nuclear density  makes the theoretical framework more microscopic and consistent, producing OPs that are quite successful in the description of the available data.  The main limitation of the NCSM is that, due to the prohibitive scaling of this approach for heavier systems, it can be used only for nuclei with $A$ not greater than 16, while in general and, in particular, for the study of nuclei away from stability, microscopic OPs are required for a wider range of nuclei. It is therefore necessary to resort to many-body approaches with better scaling with respect to the mass number that allow reaching medium-mass and heavy nuclear targets. 

We have begun exploiting the self-consistent Green's function (SCGF) theory \cite{Vorabbi8}, which presents better scaling of computational requirements with respect to the mass number and allows us to reach heavier systems, currently up to $ A\simeq 140$, providing fully nonlocal densities for the target. The density matrix has been computed using the SCGF approach and its ADC($n$) algebraic diagrammatic construction truncation scheme at different orders $n$. The standard Dyson formulation of SCGF has been used for closed-shell nuclei and its Gorkov extension for semi-magic open shells \cite{SCGF1,SCGF2,SCGF3,SCGF4,SCGF5}. The densities have been computed  with $NN$ and $3N$ chiral forces derived within the chiral effective field theory. Several chiral interactions are available,  which are able to reproduce with a high precision $NN$ phaseshifts and deuteron and triton properties. However, constraining the interactions to only few-body observables often fails to reproduce binding energies and radii of larger nuclei simultaneously with the empirical nuclear matter saturation point. Recently, it has been found that proper saturation can be recovered if light to medium mass nuclei are also used to determine the Hamiltonian \cite{NNLOSAT,PRC102}. The possibility to simultaneously account for energies
and radii of medium-mass nuclei motivated us to adopt the NNLO$_{\rm sat}$
interaction \cite{NNLOSAT} in the calculation of the densities. In particular, an accurate reproduction of the target radius is extremely important for a good description of the diffraction minima of the cross section \cite{PRL125}.

We investigated the dependence of the scattering observables on details of  \textit{ab initio} SCGF calculations and on the chiral potential used in the $NN$ $t$ matrix. Calculations are performed for Ca and Ni isotopes and the results are compared with available experimental data \cite{Vorabbi8}. Our results indicate that the SCGF input is stable and scattering observables are well converged with respect to the model space, $3N$ forces, and many-body truncation already at the ADC(2) level.

\begin{figure}[ht]
\begin{center}
\includegraphics[width=6.cm]{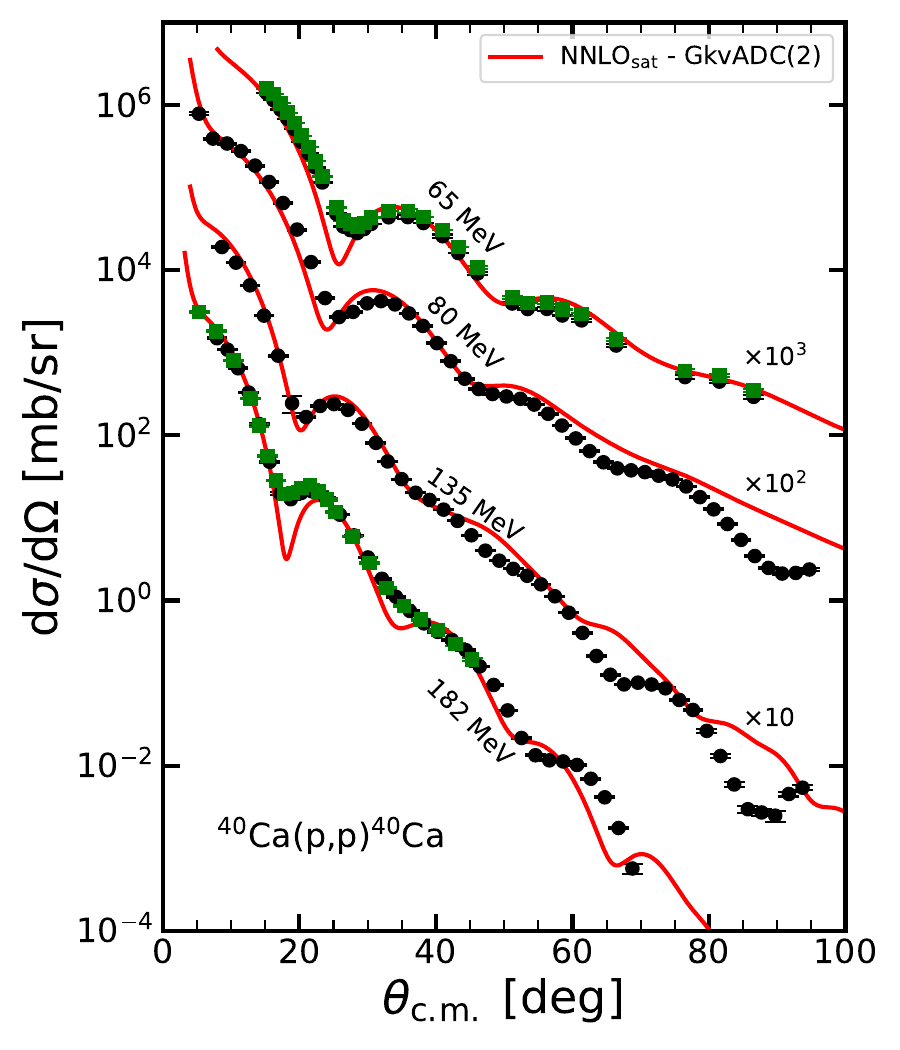}
\caption{Differential cross section as a function of the c.m.
scattering angle for elastic proton scattering off $^{40}$Ca at 65, 80,
135, and 182 MeV laboratory energies. The results of the OPs obtained using GkvADC(2) SCGF densities computed from the NNLO$_{\rm sat}$ chiral interaction
are compared to experimental data from Refs. \cite{PhysRevC.26,NPA366,PhysRevC.23a,PhysRevC.26a,AF19}. Figure from Ref. \cite{Vorabbi8}.}
\label{fig:Ca40}
\end{center}
\end{figure}

Our OPs give a good description of the experimental differential cross sections. An example is given in Figure~\ref{fig:Ca40}, where the experimental cross sections for elastic  proton scattering off $^{40}$Ca at 65, 80, 135, and 182 MeV are compared with the results of our OPs obtained using the Gorkov SCGF at second order, GkvADC(2), and with the NNLO$_{\rm sat}$ interaction. Our OPs  describe the experimental data at all energies considered, in particular, we notice the remarkable agreement at 65 MeV, an energy that can be considered at the limit of validity of the IA adopted in our OP model. The agreement with data gets somewhat worse, as usual, for larger values of the scattering angle.

\begin{figure}[ht]
\begin{minipage}[h]{0.5\textwidth}
\includegraphics[width=5.2cm]{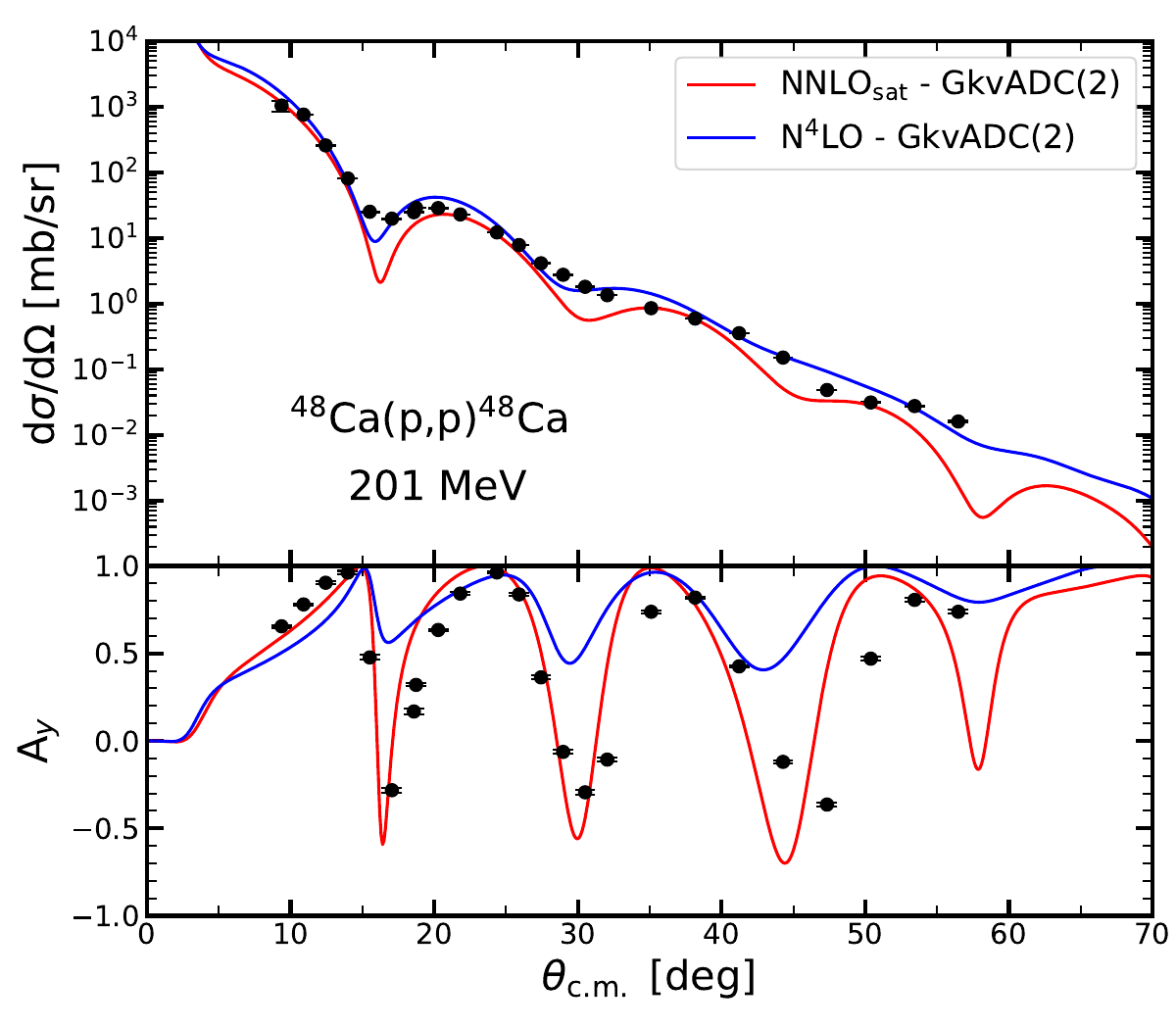}
\end{minipage} \hspace{0.01\textwidth}
\begin{minipage}[h]{0.5\textwidth}
\includegraphics[width=5.2cm]{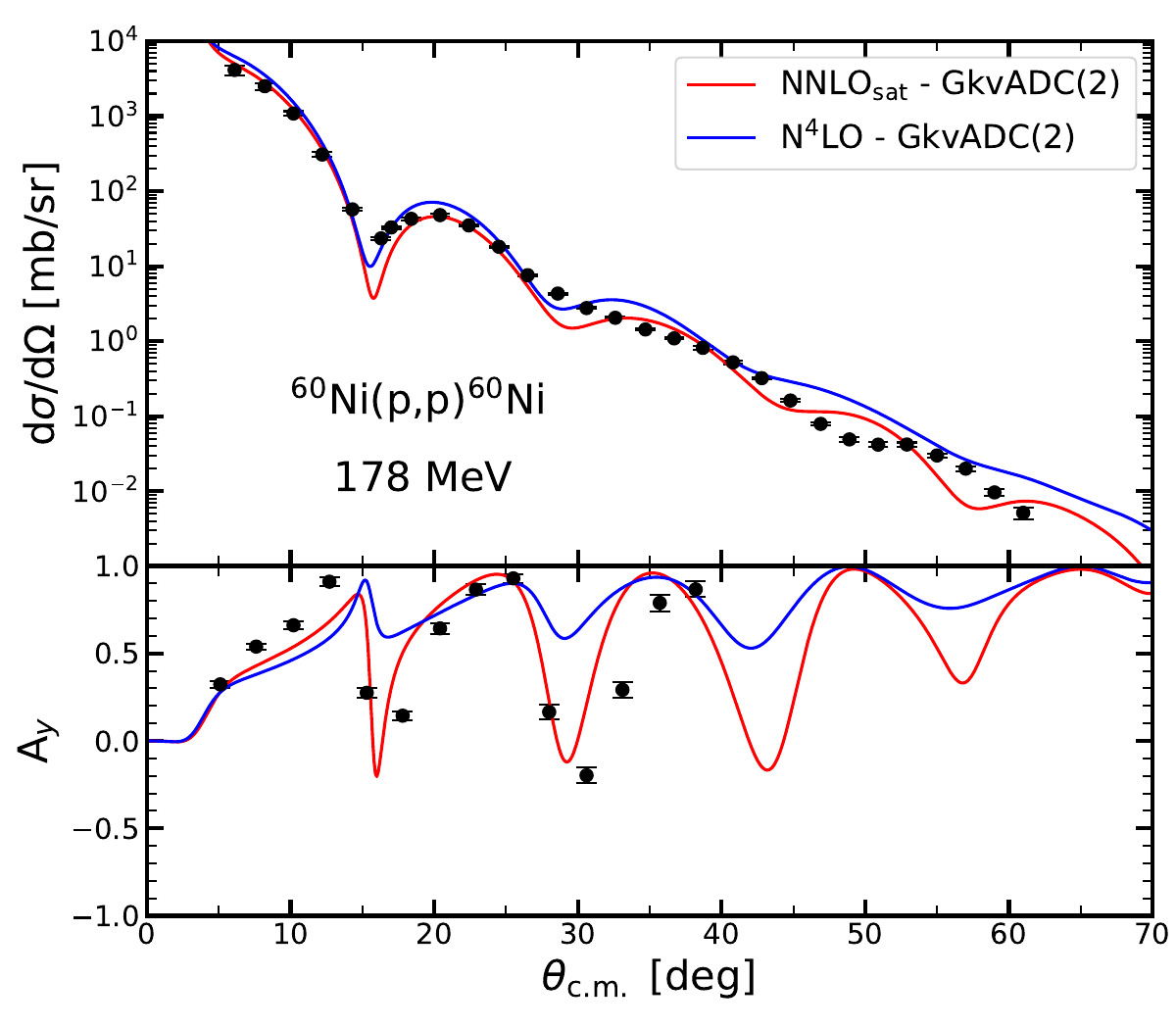}
\end{minipage}
\caption{Differential cross section (top panels) and analyzing power (bottom panels) as a function of the c.m. scattering angle for elastic proton scattering off $^{48}$Ca at 201 MeV (left panels) and $^{58}$Ni at 178 MeV 
(right panels) laboratory energy. The results of the OPs obtained using NNLO$_{\rm sat}$(red lines) and N$^4$LO (blue lines) in $t_{NN}$ are compared. The densities are obtained from GkvADC(2) SCGF calculations computed with NNLO$_{\rm sat}$. Experimental data from Refs. \cite{PhysRevC.49} (left panels) and \cite{NPA365} (right panels). Figure from Ref. \cite{Vorabbi8}.}
\label{fig:CaNi}
\end{figure}

Figure~\ref{fig:CaNi} displays the differential cross section and analyzing power as a function of the c.m. scattering angle for protons off $^{48}$Ca at 201 MeV and  $^{58}$Ni at 178 MeV. The experimental data are compared
with the results obtained using NNLO$_{\rm sat}$ and N$^4$LO \cite{EMN}
chiral interactions in  $t_{NN}$. The two interactions produce significant differences in both shape and size of the cross section and analyzing power.
Both results give a reasonable description of the experimental cross
section, although the agreement is generally better with NNLO$_{\rm sat}$. Larger differences are found for $A_y$, where both interactions describe the shape and the position of the experimental minima, but only NNLO$_{\rm sat}$ gives a remarkably good description of their depth. More results, for different isotopes and at different proton energies, confirm these findings \cite{Vorabbi8}.

Overall, the agreement found between our results and the experimental data is remarkably good and makes our approach to the OP comparable to the other existing approaches on the market. We note that the $NN$ and $3N$ chiral interactions are the only input in the calculation of our microscopic OPs. 

\section{Conclusions and perspectives}

Few years ago we started a project to obtain microscopic OPs for elastic (anti)nu\-cle\-on-nucleus scattering within the framework of  chiral effective field theories. Our OPs were derived at the first-order term of the spectator expansion of the Watson multiple-scattering theory and adopting the IA. They are obtained as a folding integral of the target density and the $NN$ $t$ matrix. The results of our OPs are in reasonably good agreement with the experimental data, for both elastic proton and antiproton-nucleus scattering.

In this contribution we have reported recent achievements of our project.

When the OP is computed with a nonlocal density from the \textit{ab initio} NCSM, $NN$ and $3N$ interactions are consistently included in the structure part. The exact treatment of the $3N$ force in the dynamic part involves multiple scattering that would make the calculation too difficult. The impact of genuine $3N$ forces has been evaluated, already at the level of the  single-scattering approximation, averaging them over the Fermi sphere and thus defining a density-dependent $NN$ interaction which acts as a medium correction of the bare $NN$ potential and which is then added to the bare $NN$ potential in the calculations of the  $NN$ $t$ matrix. The effect of this $3N$ force is generally very small on the cross sections but can be sizeable on polarization observables.  Of course a more complete treatment of $3N$ forces would be required.

The extension of our microscopic OPs to nonzero spin targets provides a good description of the data, of the same quality as the one obtained for zero spin targets, and allows us to give reliable predictions for a wider range of  stable and unstable nuclei.

The use of \textit{ab initio} densities from the SCGF theory allows us to extend our OP to heavier targets. The combination of the spectator model and SCGF theories offers good opportunities for the physics of radioactive beams, in particular, toward the solution of the long-standing issue of the lack of consistency between structure and reactions in the interpretation of data.

The SCGF theory can provide two-nucleon spectral densities \cite{2NSF}, which are the basis for extending the OP model to the next term of the spectator expansion. At energies where the IA may become questionable, the self-energy computed through SCGF theory is itself a viable \textit{ab initio} OP \cite{Idini}. Future work in these directions would allow us to extend the energy range of applicability of the microscopic OP.

\section*{Acknowledgements}
The work reported in this contribution has been obtained in collaboration with
C. Barbieri, M. Gennari, R. Machleidt, P. Navr{\'a}til, and V. Som\`a. We thank all of them.

\end{document}